\documentclass{mn2e}
\newcommand{\kms}{\ifmmode {\rm km\ s}^{-1} \else km s$^{-1}$\ \fi}
\newcommand{\ergs}{\ifmmode {\rm erg\ s}^{-1} \else erg s$^{-1}$\ \fi}

\newcommand{\feii}{Fe {\sc ii}\ }
\newcommand{\mgii}{Mg {\sc ii}\ }
\newcommand{\civ}{C {\sc iv}\ }

\newcommand{\lb}{\ifmmode L_{\rm Bol} \else $L_{\rm Bol}$\ \fi}
\newcommand{\ledd}{\ifmmode L_{\rm Edd} \else $L_{\rm Edd}$\ \fi}
\newcommand{\lx}{\ifmmode L_{\rm 2-10keV} \else  $L_{\rm 2-10keV}$\ \fi}
\newcommand{\hb}{\ifmmode H\beta \else H$\beta$\ \fi}
\newcommand{\ha}{\ifmmode H\alpha \else H$\alpha$\ \fi}
\newcommand{\oiii}{[O {\sc iii}]\ }
\newcommand{\oii}{[O {\sc ii}]\ }
\newcommand{\nii}{[N {\sc ii}]\ }
\newcommand{\sii}{[S {\sc ii}]\ }
\newcommand{\mbh}{\ifmmode M_{\rm BH}  \else $M_{\rm BH}$\ \fi}
\newcommand{\lv}{\ifmmode \lambda L_{\lambda}(5100\AA) \else $\lambda L_{\lambda}(5100\AA)$\ \fi}
\newcommand{\msun}{M_{\odot}}
\newcommand{\mdot}{\ifmmode \dot{m} \else \dot{m} \fi }
\newcommand{\llog}{\ifmmode {\rm log} \else {\rm log} \fi }

\usepackage{graphicx}
\usepackage{color}
\usepackage{lscape}

\begin{document}
\title[Mass in NLS1s]{A note on black hole masses estimated by the second moment in Narrow Line Seyfert 1 Galaxies}
\author[Bian et al.]
{Wei-Hao Bian$^{1,2}$, Chen Hu$^{1}$, Qiu-Sheng Gu$^{3}$, Jian-Min Wang$^{1,4}$  \\
$^{1}$ Key Laboratory for Particle Astrophysics, Institute of High
Energy Physics, Chinese Academy of Sciences, Beijing 100039,
China\\
$^{2}$ Department of Physics and Institute of Theoretical Physics,
Nanjing Normal University, Nanjing
210097, China\\
$^{3}$ Department of Astronomy, Nanjing University, Nanjing
210093, China\\
$^4$  Theoretical Physics Center for Science Facilities, Chinese
Academy of Sciences, China \\} \maketitle

\begin{abstract}
The second moment of the H$\beta$ emission line is calculated for
329 narrow line Seyfert 1 galaxies (NLS1s) selected from the Sloan
Digital Sky Survey (SDSS), which is used to calculate the central
supermassive black hole (SMBHs) mass of each. We find that the
second moment depends strongly on the broader component of the
H$\beta$ line profile. We find that for the NLS1s requiring two
Gaussians to fit the H$\beta$ line the mean value of the SMBH mass
from the H$\beta$ second moment is larger by about 0.50 dex than
that from the full width at half maximum (FWHM). Using the gas
velocity dispersion of the core/narrow component of \oiii $\lambda$
5007 to estimate the stellar velocity dispersion, $\sigma_{*}$, the
new mass makes NLS1s fall very close to the $\mbh - \sigma_{*}$
relation for normal AGNs. By using $\sigma_{*}$  measured directly
from SDSS spectra with a simple stellar population synthesis method,
we find that for NLS1s with mass lower than $10^7 \msun$, they fall
only marginally below the $\mbh - \sigma_{*}$ relation considering
the large scatter in the mass calculation.

\end{abstract}


\section{INTRODUCTION}
Narrow line Seyfert 1 galaxies (NLS1s) are thought to be a special
subclass of active galactic nuclei (AGNs) harboring relatively small
but growing supermassive black holes (\mbh, SMBHs), compared to
other broad line Seyfert 1 galaxies (BLS1s; e.g., Osterbrock \&
Pogge 1985; Boller et al. 1996; Mathur et al. 2000). Whether NLS1s
follow the well-known $\mbh - \sigma_{*}$ (or $\mbh - L_{\rm
bulge}$) relation defined in inactive galaxies is a question open to
debate, where the stellar velocity dispersion ($\sigma_{*}$) is
measured at an eighth of the effective radius of the galaxies (e.g.,
Tremaine et al. 2002; Mathur Kuraszkiewicz \& Czerney 2001; Bian \&
Zhao 2004; Grupe \& Mathur 2004; Botte et al. 2005; Barth et al.
2005; Watson Mathur \& Grupe 2007; Ryan et al. 2007;  Komossa \& Xu
2007). It remains also a question for other types of AGNs (e.g.,
Nelson et al. 2000; Greene \& Ho 2006; Shen et al. 2008; Woo et al.
2008). In investigating the $\mbh - \sigma_{*}$ relation for NLS1s
or/and other AGNs, the method of determinging $\mbh$ and
$\sigma_{*}$ is very important.

The central SMBH mass in AGNs is a key parameter to understand the
nuclear energy mechanism as well as the cosmic formation and evolution of
SMBHs and their host galaxies (e.g. Rees 1984; Gebhardt et al. 2000;
Ferrarese \& Merrit 2000; Tremaine et al. 2002). In the past two
decades, there has been striking progress in finding more reliable
methods to calculate SMBHs masses in AGN through the line width, $\Delta V$, of
H$\beta$ (or H$\alpha$, \mgii, \civ) from the broad line region
(BLR) and the BLR size, $R_{\rm BLR}$ (e.g., Kaspi et al. 2000;
McLure \& Dunlop 2004; Bian \& Zhao 2004; Peterson et al. 2004;
Greene \& Ho 2005b).

Introducing the scaling factor, $f$, to characterize the unclear
kinematics and the geometry of the BLRs, the SMBH mass is
calculated by:
\begin{equation}
M_{\rm BH} = f\frac{R_{\rm BLR} \Delta V^2}{G}.
\end{equation}
The uncertainties of SMBHs masses in AGN from Equation 1 are
mainly from the uncertainties in $R_{\rm BLR}$,  $f$ and $\Delta
V$.

Much effort has been focused on determining $R_{\rm BLR}$, from the
reverberation mapping method or empirical size-luminosity relations
(e.g., Kaspi et al. 2000; 2005; Vestergaard \& Peterson 2006; Bentz
et al. 2006). There are mainly two ways to parameterize the line
widths of broad emission lines, i.e., the full width at half maximum
(FWHM) and the second moment ($\sigma_{line}$). For a Gaussian line
profile, $\rm FWHM/\sigma_{\rm line}=\sqrt{8ln2} \approx 2.35$;
while for a Lorentzian profile, $\sigma_{\rm line} \rightarrow
\infty$. However, people usually find that one Gaussian component
provides a poor fit to the H$\beta$ emission line profile after
subtracting the contribution from narrow line region (NLR),
especially for NLS1s (e.g., Rodriguez-Ardila et al. 2000; Dietrich
et al. 2005; Mullaney \& Ward 2008). Salviander et al. (2007) used a
Gauss-Hermite function to measure the H$\beta$ FWHM (also see McGill
et al. 2008). Netzer \& Trakhtenbrot (2007) measure FWHM from the
two-Gaussian fits to the H$\beta$ profile (see also Mullaney \& Ward
2008). For 12 NLS1s, Dietrich et al. (2005) suggested that their
broad emission line profiles are well represented by employing two
Gaussian components, while the Lorentzian profile does not do as
well for the core and wing of the H$\beta$ profile simultaneously.
Based on the analysis of reverberation mapping data, it was
suggested that $\sigma_{line}$ rather than FWHM be used to
characterize the line width (Fromerth \& Melia 2000; Krolik et al.
2001; Peterson et al. 2004).

For 16 AGNs with BLRs sizes from the reverberation mapping and possessing
reliable $\sigma_{*}$ measurements, Onken et al. (2004) determined the scaling
factor, $f$, to make the reverberation-based SMBH masses
consistent with the well-known $\mbh - \sigma_{*}$ relation
of inactive galaxies (Tremaine et al. 2002). They found
that $f$ is $5.5\pm 1.8$ when $\sigma_{\rm line}$ from
root-mean-square (rms) spectra is adopted as $\Delta V$. Collin et
al. (2006) proposed different $f$ factors for emission lines with different
ratios of FWHM to $\sigma_{\rm line}$. They suggested $f=3.85\pm
1.15$ when $\sigma_{\rm line}$ from mean spectra is adopted as
$\Delta V$. It is also often assumed that the BLR gas has random orbits.
Netzer (1990) suggested that $f=3$ when FWHM/2 is adopted
as $\Delta V$ (e.g., Kaspi et al. 2000; 2005; Greene \& Ho 2006).

Although $\sigma_{*}$ is difficult to measure for AGNs because the
nuclei outshine their hosts, there are a larger number of AGNs from
the Sloan Digital Sky Survey (SDSS) with obvious stellar absorption
features within 3'' aperture spectra, which can be used to measure
$\sigma_{*}$ (e.g., Green \& Ho 2006, Shen et al. 2008). About the
method to measure $\sigma_{*}$, please refer to Bian et al. (2007)
and the reference therein. The gaseous velocity dispersion (e.g.,
$\sigma_{\rm [O~III]}^{core}$, $\sigma_{\rm [N~II]}$) is often used
to as a proxy for $\sigma_{*}$ (e.g., Nelson \& whittle 1996; Greene
\& Ho 2005a).

In our previous work based on NLS1s selected from the Sloan Digital
Sky Survey early data release (SDSS EDR), using H$\beta$ FWHM to
calculate the mass and \oiii narrow/core FWHM to give the
$\sigma_{*}$, we found that the SMBH masses of NLS1s deviated
significantly from the well-known $\mbh - \sigma_{*}$ relation (Bian
\& Zhao 2004, Bian et al. 2006). Here, we use the second moment of
the broad H$\beta$ profile to re-investigate SMBHs masses in NLS1s.
All of the cosmological calculations in this paper assume $\rm
H_{\rm 0}=70~ \kms \rm ~Mpc^{-1}$, $\Omega_{\rm M}=0.3$, and
$\Omega_{\Lambda} = 0.7$.

\section{Sample and Data Analysis}
\begin{figure*}
\begin{center}
\includegraphics[width=12cm]{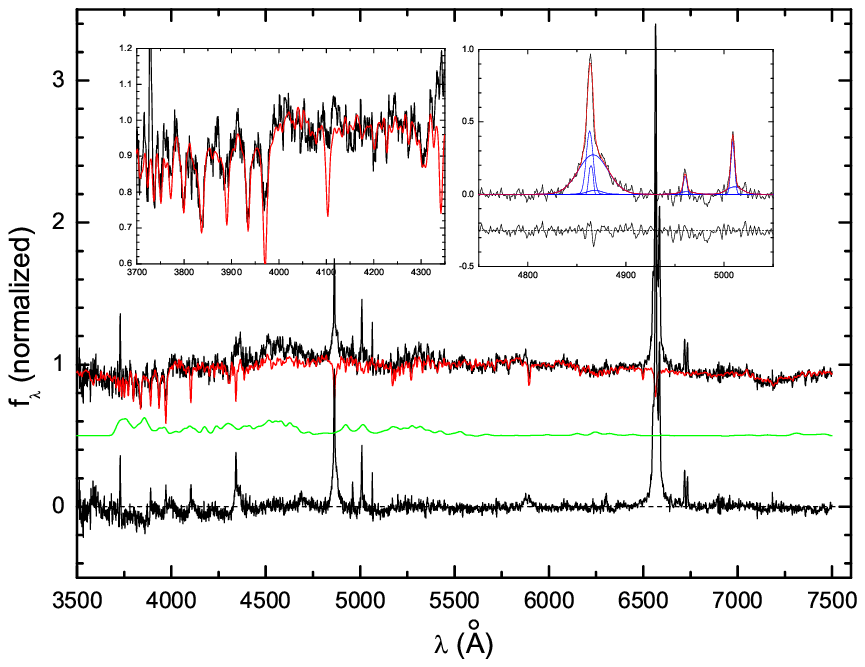}
\includegraphics[width=12cm]{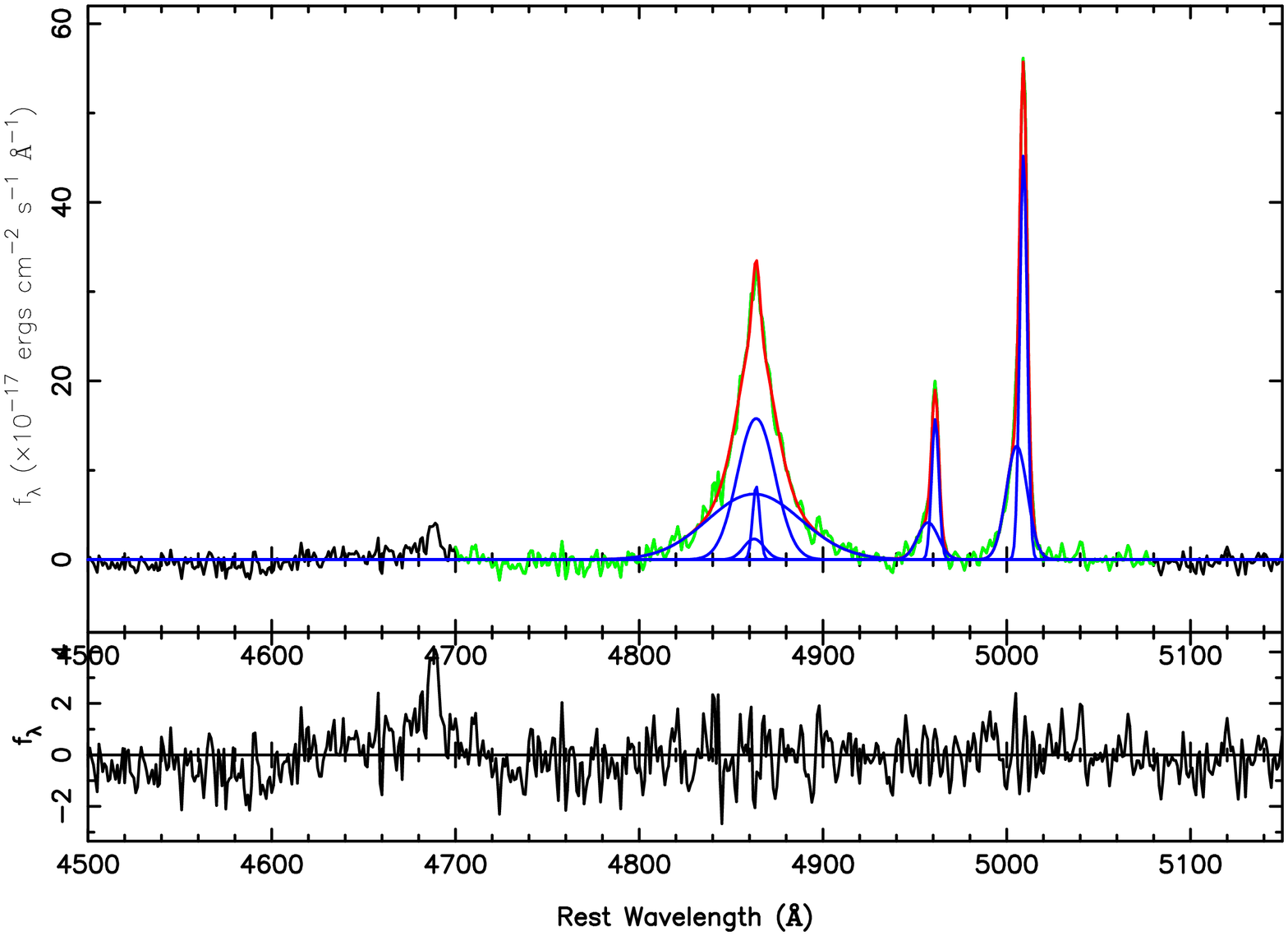}
\caption{$Top$: An example fit for SDSS J123831.34+644356.5. The
black line is the original spectrum after Galactic-extinction
correction in the rest frame. The red line is the contribution from
the host galaxy. The green line is the \feii emission. The residual
is shown at the bottom. The left upper panel shows the region around
Ca H+K $\lambda \lambda$ 3969, 3934 and G-band. In the right upper
panel we show the H$\beta$ and \oiii line fits, where the blue one
with the highest peak is the IC, the blue one with the second
highest peak is the BC, the two blue ones with the lower peaks are
the components from NLRs. $Bottom$: An another example of line fit
for SDSS J084716.88+334858.9. The black line is the original
spectrum after Galactic-extinction correction, the starlight
subtraction, and the \feii subtraction in the rest frame. The bottom
panel is the residual. The multiple Gaussian components are in blue
and the sum of them is in red. The fitting window is in green.}
\end{center}
\end{figure*}

We use a largest sample of about 2000 NLS1s from SDSS DR3 (Zhou et
al. 2006). Zhou et al. (2006) presented this sample of about 2000
NLS1s selected from objects assigned as "QSOs" and "galaxies" in the
spectroscopic database of SDSS DR3. The only criterion is that the
broad component of \hb or \ha is detected and is narrower than 2200
\kms in FWHM. Using a Lorentzian function to model the broad
H$\beta$ profile, Zhou et al. (2006) obtained the FWHM and
calculated SMBHs masses from the $R_{\rm BLR} - L_{5100}$ relation
of Kaspi et al. (2000). We use the latest SDSS spectra from data
release 6 (DR6).

We briefly outline our steps to do the SDSS spectral analysis. (1)
We use simple stellar population (SSP) synthesis to model the
stellar contribution in the Galactic extinction-corrected spectra in
the rest frame (Cid Fernandes et al. 2005; Bian et al. 2007). We
include 45 templates from Bruzual \& Charlot (2003; BC03) and one
power-law component (representing the AGN continuum emission).
During the stellar population synthesis, we put twice the weight in
the fit for strongest stellar absorption features, such as CaII K,
G-band, and Ca II$\lambda \lambda$ 8498, 8542, 8662 triplet. For
details, please see Bian et al. (2007) and the references therein.
(2) The optical and ultraviolet \feii template from the prototype
NLS1 I ZW 1 is used to subtract the \feii emission from the residual
spectra after the above step. (3) Four Gaussians are used to model
the H$\beta$ profile, and two sets of two Gaussians are used to
model the \oiii$\lambda \lambda 4959, 5007$ lines. We take the same
linewidth for each component, and fix the flux ratio of
\oiii$\lambda$4959 to \oiii $\lambda$5007 to be 1:3. Two components
of \hb (from the NLR) are set to have the same linewidth of each
component of \oiii $\lambda$5007 and their flux are constrained to
be less than 1/2 of each component of \oiii $\lambda 5007$. Two
broad components are used to model broad H$\beta$ profile from the
BLR contribution (broad component and intermediate component, BC and
IC). We present our spectral fitting method in detail in Hu et al.
(2008). In the above three steps, the best fit is reached by
minimizing $\chi^2$, $\chi^2=\sum_{i} \left(\frac{y_i-y_{\rm
model}}{\sigma_i}\right)^2/N $, where $\sigma_i$ is the error of the
data set $(x_i,y_i)$, N is the degrees of freedom.

Objects without clear \hb or \oiii lines are eliminated. In order to
obtain a reliable spectral fit, we carefully select objects for
analysis. We select objects by the following criteria: (1) the
signal-to-noise ratio (S/N) is larger than 15. The S/N is measured
in the wavelength 4800-5040\AA, covering the range for line-fitting.
(2) $\chi^2$ in above three steps (SSP, \feii subtraction, and \hb
and \oiii lines fitting) are less than 2.5. (3) the FWHM errors of
BC and IC of \hb from BLR and \oiii$\lambda \lambda 4959, 5007$ are
less than $100\%$. (4) The height of one component of \oiii is less
than the height of one component of \oiii by 30\%. The first
criterion leads to about 900 NLS1s. The second and third criteria
make sure we have a good fitting of H$\beta$ and \oiii. The fourth
criterion make us reliable to obtain the core/narrow component of
\oiii line, and their line width is used to trace the $\sigma_{*}$.
Then we visually check these spectra one by one. In the end, we have
a sample of 329 NLS1s.

The first moment of the line profile is
\begin{equation}
\lambda_{0}  = \frac{\int \lambda P(\lambda) d\lambda}{\int
P(\lambda) d\lambda},
\end{equation}
The second moment of the line profile is
\begin{equation}
\sigma^2_{line}=\frac{\int \lambda^2 P(\lambda) d\lambda}{\int
P(\lambda) d\lambda}-\lambda_{0}^2.
\end{equation}
We use the BC and IC to reconstruct the broad H$\beta$ profile as
$P(\lambda)$. Then we use the method of Peterson et al. (2004) to
measure the FWHM and $\sigma_{\rm H\beta}$ from the reconstructed
broad H$\beta$ profile (Equations (2-3)). The error of $\sigma_{\rm
H\beta}$ is calculated from the error of FWHM for BC and IC in the
line fit. The FWHM measurement is not sensitive to the broad wing of
H$\beta$, while the measurement of H$\beta$ second moment is. We
have to make sure the necessary of using BC and IC in the H$\beta$
fitting (see also Dietrich et al. 2005). We use one broad component
to model the H$\beta$ profile from the BLR at the same time. When
the $\chi^2$ in the fitting is decreased by 20\% with two-Gaussian
respect to that with one-Gaussian, We have to make sure it is
necessary to use BC and IC, and we use the result of two-Gaussian
fitting, otherwise we use the result of one-Gaussian fitting. For
H$\beta$ profile with one-Gaussian fitting, the second moment is
directly from our fitting, not from Equations (2-3). For all
objects, the measurement of FWHM is adopted from the two-Gaussian
reconstructive profile. Hereafter, we call the total 329 NLS1s as
sample A, 209 NLS1s with two-Gaussian fitting as sample B, and the
other 120 NLS1s with one-Gaussian fitting as sample C.

As the strongest forbidden line in the SDSS spectral wavelength
coverage, we use the gas velocity dispersion of the narrow/core
\oiii component from NLRs to trace $\sigma_{*}$,
$\sigma^{core}_{\rm[O III]}=\sqrt{\sigma_{\rm obs}^2-[\sigma_{\rm
inst}/(1+z)]^2}$, where $z$ is the redshift. For SDSS spectra, the
mean value of instrumental resolution $\sigma_{\rm inst}$ is 60
\kms\ for \oiii (e.g., Greene \& Ho 2005a). We also reliably obtain
$\sigma_{*}$ (correction the resolutions of the SDSS spectra and
BC03 templates) from our SSP synthesis for about 98 NLS1s. The error
of $\sigma_{*}$ is given by different typical errors for different
effective S/Ns at 4020 \AA\ (Bian et al. 2007).

\section{Mass from $\sigma_{\rm H\beta}$ and FWHM}
\subsection{Mass estimation}
\begin{figure*}
\begin{center}
\includegraphics[width=16cm]{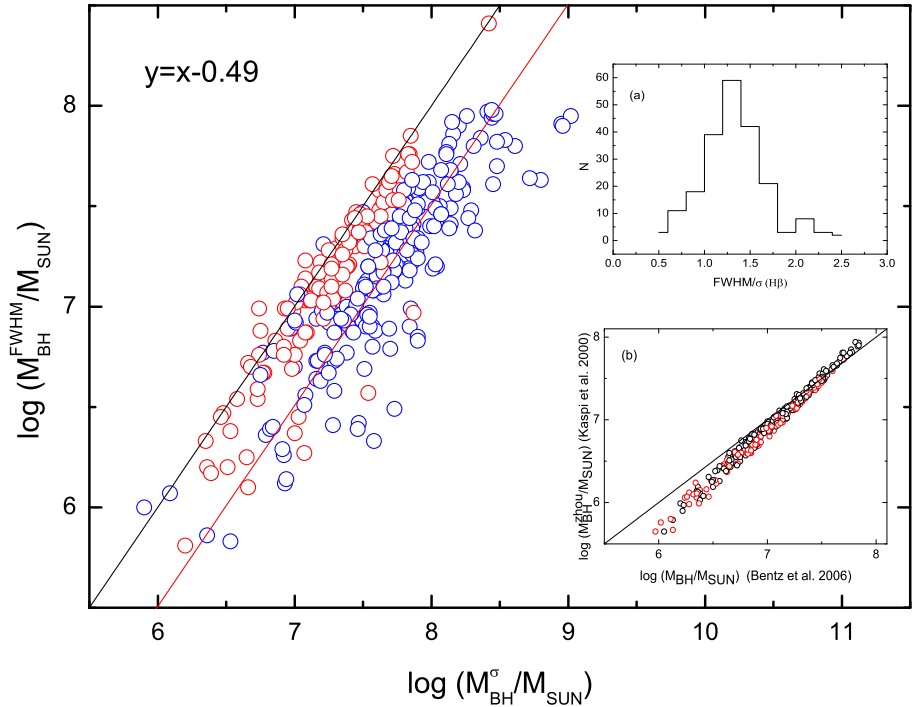}
\caption{The mass from the H$\beta$  FWHM versus the mass from the
H$\beta$ second moment. The blue circles denote objects in sample B
where the H$\beta$ from BLRs can be fitted well by two-Gaussian with
$\chi^2$ decreased by at least 20\% with respect to that by one
Gaussian. The red circles denote objects in sample C with the
H$\beta$ from BLRs can be fitted well by one Gaussian in order to
avoid the broad wing effect in measurement of the second moment. The
solid line is 1:1, and the red line is the best linear fit with the
fixed slope of 1 for sample B, y=x-0.49. Then FWHM and $\sigma_{\rm
H\beta}$ are measured from the reconstructed broad H$\beta$ profile.
For sample C, there is a scatter from 1:1. For sample C, mass from
$\sigma_{\rm line}$ can be increased by 0.13 dex with respect to
that from FWHM. (a) The distribution of $\rm FWHM/\sigma (H\beta)$
for 209 NLS1s in sample B. (b) SMBH mass from Zhou et al. (2006)
versus that with updated the $R_{\rm BLR}-L_{5100}$ relation of
Bentz et al. (2006).}
\end{center}
\end{figure*}

With our measurement of $\sigma_{\rm H\beta}$, we use the more
recent $R_{\rm BLR} - L_{5100}$ relation (which seems to hold for
NLS1s; Peterson et al. 2004) of Bentz et al. (2006) and $f=3.85$
(Collin et al. 2006) to calculate the SMBHs masses in NLS1s. We also
calculate the mass from our FWHM for the H$\beta$ profile from the
BLR by $R_{\rm BLR} - L_{5100}$ relation of Bentz et al. (2006) and
$f=3$ (Netzer 1990). In Figure 2, we compare these two masses. For
sample B, the best linear fit with fixed slope of 1 gives y=x-0.49,
the mass from $\sigma_{\rm H\beta}$ is on average larger by 0.49 dex
than that from FWHM. In Figure 2a, we show the distribution of $\rm
FWHM/\sigma$, $1.33\pm 0.36$, which deviates from 2.35 for a
Gaussian profile. Our mass correction from $\sigma_{\rm H\beta}$
respective to mass from FWHM is mainly due to the H$\beta$
emission-line profile deviation from the Gaussian profile. For six
NLS1s in Peterson et al. (2004), we find that the mass based on
$\sigma_{\rm H\beta}$ is on average larger by 0.46 dex with respect
to that from FWHM, which is consistent with our calculation.

%It is possible that objects in sample B are BLS1s, although it is
%popularly reported that the H$\beta$ emission line in NLS1s is
%better fit with a Lorentzian profile.

By using the FWHM derived from the Lorentzian profile, Zhou et al.
(2006) calculated SMBHs masses from the $R_{\rm BLR} - L_{5100}$
relation of Kaspi et al. (2000, $R_{\rm BLR} \propto
L_{5100}^{0.7}$) and $f=3$.  For comparison with the results of
Zhou et al. (2006), we use the updated $R_{\rm BLR} - L_{5100}$
relation of Bentz et al. (2006, $R_{\rm BLR} \propto
L_{5100}^{0.518}$) and $f=3$ to calculate the mass. For the FWHM
from Zhou et al. (2006), the updated $R_{\rm BLR} - L_{5100}$
relation of Bentz et al. (2006) would lead the masses of Zhou et al.
(2006) to be larger by 0.1-0.3 dex for NLS1s with mass less than $10^7
\msun$ (see Figure 2b). By the best linear fit through zero,
we find FWHM in Zhou et al. (2006) derived from the Lorentzian
profile is $0.84 \pm 0.01$ of our FWHM for objects in sample B,
which leads to the SMBH mass decreasing by 0.15 dex if we use the FWHM
of Zhou et al. (2006).

The usa of $R_{\rm BLR}-L_{5100}$ relation of Kaspi et al.
(2005) will decrease $\mbh$ by 0.17 dex with respect to that of Kaspi
et al. (2000). If we use $f=3.85$ instead of $3$, the mass will
increase by 0.11 dex. The uncertainty of the mass calculation
from the H$\beta$ line is mainly from the systematic uncertainties, up
to about 0.5 dex, which is due to the unknown kinematics and geometry
in BLRs, and perhaps the effects of radiation pressure
(e.g., Krolik 2001; Peterson et al. 2004;
Decarli et al. 2008; Marconi et al. 2008).

\subsection{The $\mbh - \sigma^{\rm core}_{\rm [O~III]}$ relation}
\begin{figure*}
\begin{center}
\includegraphics[width=16cm]{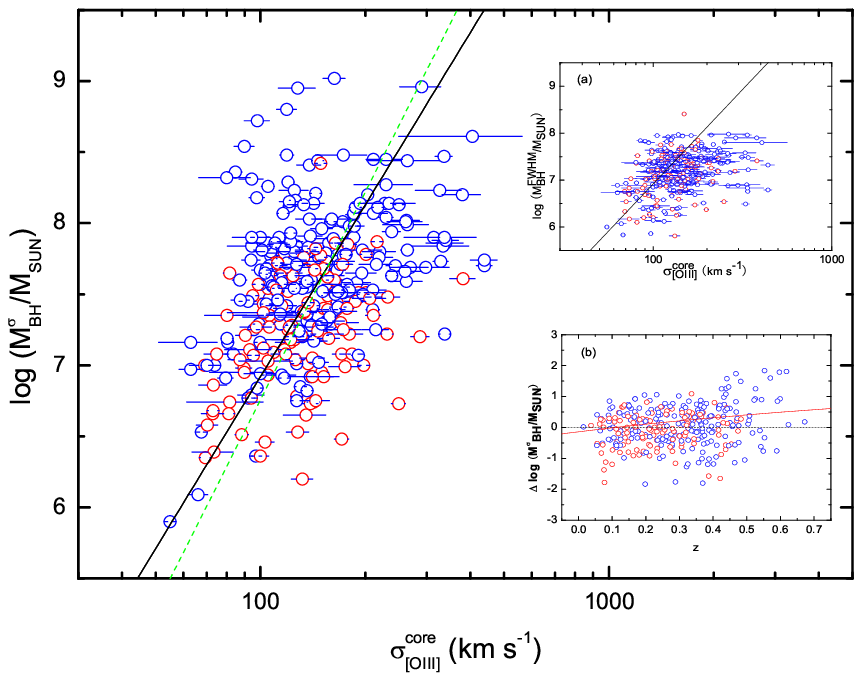}
\caption{The SMBH mass versus the core/narrow \oiii gas velocity
dispersion. The mass is derived from the H$\beta$ second moment.
The sybols are same as Figure 2. The solid line is the $\mbh -
\sigma_{*}$ relation of Tremaine et al. (2002). The green dashed
line is the $\mbh - \sigma_{*}$ relation of Ferrarese \& Ford
(2005). (a) SMBH mass from H$\beta$ FWHM versus the \oiii gas
velocity dispersion. (b) Mass deviation versus the redshift. The
solid line is the relation of Woo et al. (2008).}
\end{center}
\end{figure*}

In Figure 3, we show the SMBH mass from $\sigma_{\rm H\beta}$ versus
$\sigma_{\rm [O III]}^{core}$. In Figure 3a, we show the mass
derived from our FWHM versus $\sigma_{\rm [O III]}^{core}$. It is
obvious that the mass from FWHM deviates from Tremaine et al.
relation (solid line in Figure 3a), which is consistent with the
result of Zhou et al. (2006). In Figure 3a, we show the SMBH mass
from $\sigma_{\rm H\beta}$ versus that from our FWHM. The SMBH
masses based on $\sigma_{\rm H\beta}$ in 209 NLS1s of sample B is
larger by about 0.5 dex with respect to that from FWHM.

For the solid line in Figure 3, the $\mbh - \sigma_{*}$ relation,
$\mbh (\sigma_*) = 10^{8.13}[\sigma_{*}/(200 \ \rm km
s^{-1})]^{4.02} ~~\msun $ (Tremaine et al. 2002), we calculate the
deviation of mass (see Table 1). In Figure 3b, we show the mass
deviation from the Tremaine et al. relation versus the redshift. The
dashed line is for no deviation, and the solid line is the relation
found by Woo et al. (2008).  For redshifts of NLS1s larger than 0.4,
their mean mass deviation is $0.28\pm 0.82$ respect to Tremaine et
al. relation, and for NLS1s with redshift less than 0.4, the mean
mass deviation is $-0.04\pm 0.58$ (Woo et al. 2008). We calculate
the Eddington ratio, i.e., the ratio of the bolometric luminosity
($L_{\rm bol}$) to the Eddington luminosity ($L_{\rm Edd}$), where
$L_{\rm Edd}=1.26 \times 10^{38} (M_{\rm BH}/ \msun) \ergs$. The
bolometric luminosity is calculated from the monochromatic
luminosity at 5100\AA\ , $L_{\rm bol}=c_{\rm B}\lv$, where we adopt
the correction factor $c_{\rm B}$ of 9 (Kaspi et al. 2000; Richards
et al. 2006; Netzer \& Trakhtenbrot 2007). We do not find larger
mass deviations for objects with larger Eddington ratio. In Table 1,
we give the mass, Eddington ratio, and the mass deviation from the
$\mbh - \sigma_{*}$ relation for A, B, C samples. For sample B, mass
from $\sigma_{\rm H\beta}$ can be increased by $\sim$ 0.5 dex with
respect to that from FWHM. For sample C, mass from $\sigma_{\rm
line}$ can be increased by 0.13 dex with respect to that from FWHM,
which is due to the values of $f$ that we are using for sigma ($f =
3.85$) and FWHM/2 ($f = 3$) are not exactly consistent with a single
Gaussian, whereas in order to get the same mass for a single
Gaussian the ratio of f values is 1.38. We also notice that, for
NLS1s with mass larger than $10^8 \msun$, they tend to fall above
the Tremaine et al. relation.

\begin{table}
\caption{The SMBH masses from the $\sigma_{\rm H\beta}$ and FWHM.}
\begin{tabular}{lrrrrrr} \hline \hline
& A & B & C \\
\hline
\llog $M_{\rm BH}^{\sigma}$          & $ 7.53 \pm 0.50$ & $ 7.69 \pm 0.48$ & $ 7.25 \pm 0.40$ \\
$L_{\rm bol}/L_{\rm Edd}(\sigma)$             & $-0.81 \pm 0.28$ & $-0.84 \pm 0.27$ & $ -0.75 \pm 0.28$ \\
$\Delta \llog\mbh^{\sigma}$                     & $ 0.03 \pm 0.65$ & $0.12 \pm 0.68$  & $ -0.13 \pm 0.55$\\
\hline
\llog $M_{\rm BH}^{\rm FWHM}$                      & $ 7.17 \pm 0.44$& $ 7.20 \pm 0.44$  & $ 7.12 \pm 0.43$\\
$L_{\rm bol}/L_{\rm Edd}(\rm FWHM)$           & $ -0.44\pm 0.37$ & $-0.35 \pm 0.37$ & $ -0.61 \pm 0.31$\\
$\Delta \llog \mbh^{\rm FWHM}$                       & $ -0.34\pm 0.65$ & $-0.38 \pm 0.70$ & $ -0.26 \pm 0.57$\\
\hline
\end{tabular}

\dag: A: for total 329 NLS1s; B: for 209 NLS1s in sample B; C:
for 120 NLS1s in sample C. The mass deviation is based on Tremaine et al.'s
relation when $\sigma_{\rm [O III]}^{core}$ is used for $\sigma_{*}$.
\end{table}

\section{Discussion}
\subsection{Large SMBH masses in NLS1s?}
\begin{figure}
\begin{center}
\includegraphics[width=8cm]{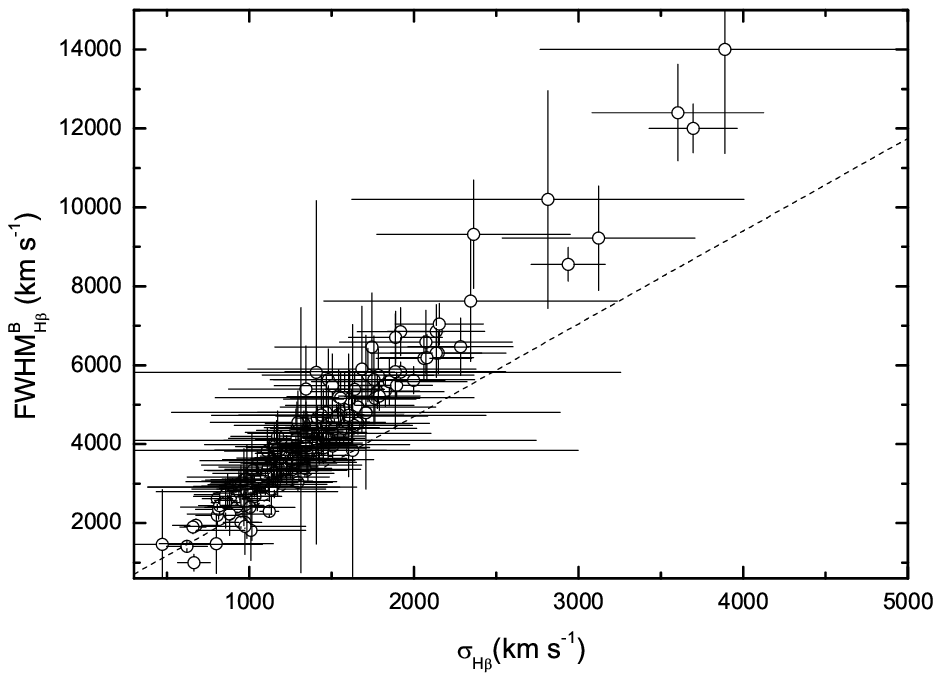}
\includegraphics[width=8cm]{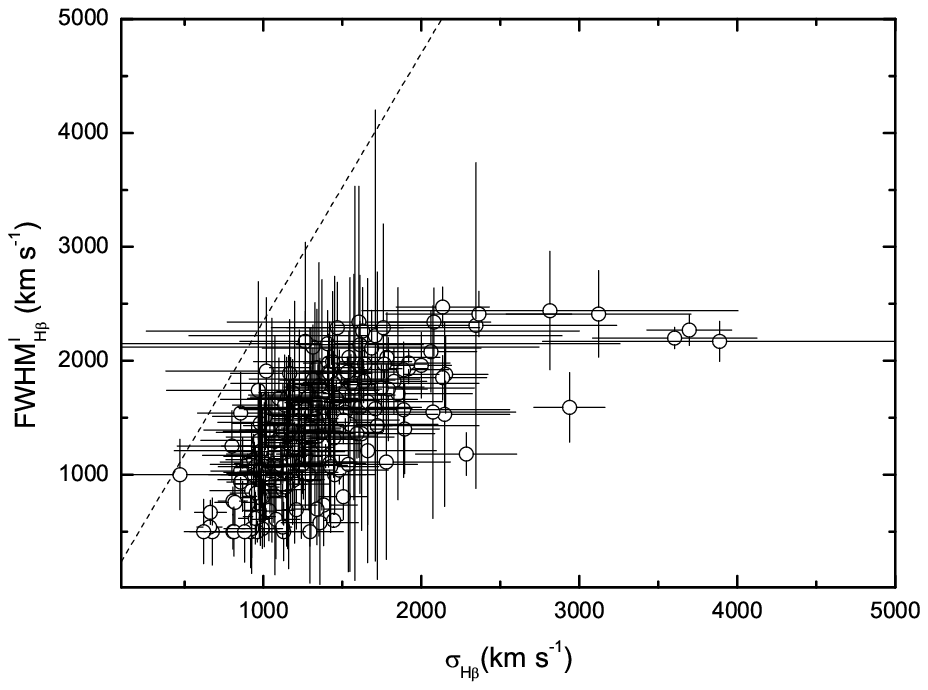}
\caption{The FWHM of BC ($top$) and IC ($bottom$) versus the
H$\beta$ second moment for sample B of 209 NLS1s. The dashed
line is the relation between FWHM and the second moment, $\rm
FWHM=\sigma_{\rm H\beta} \times 2.35$, for a Gaussian line
profile.}
\end{center}
\end{figure}

In Figure 3, we find that, for some NLS1s, mass from $\sigma_{\rm H
\beta}$ is very large, up to $10^8 - 10^9\msun$, which is due to
their obvious broad wings in their H$\beta$ profiles. With the
near-IR imaging data, Ryan et al. (2007) found that the average SMBH
mass in their NLS1s sample is $10^{7.9} \msun$, where the mass is
calculated from the mass and host galaxy luminosity relation. The
mass is typical for broad line Seyfert 1 galaxies. They find that it
is larger by 1.5 dex respect to the mass calculated from the FWHM of
Veron-Cetty et al. (2001). We note that these objects have broad
wings in their H$\beta$ profile, especially in their H$\alpha$
profile (Figure 2 in Veron-Cetty et al. 2001). It is possible that
the sample of Zhou et al. (2006) have some objects that should not
be classified as NLS1s. Considering the effects of the random
velocity and the inclination in SMBH mass estimation of NLS1s,
Decarli et al. (2008) find that these effects would increase the
mass for NLS1s by 0.84 dex, which can account for the mass
difference between NLS1s and BLS1s. The definition of NLS1s need to
be reexamined if we think that NLS1s harbor rapid growing small SMBH
with high accretion rate.

\subsection{$\sigma_{\rm H\beta}$ and FWHMs of BC, IC}
The flux ratio of H$\beta$ to \oiii $\lambda$5007 from NLRs is
often taken to be around 10\% (e.g. Osterbrock \& Pogge 1985; McGill
et al. 2008). Rodriguez-Ardila et al. (2000) suggested this ratio
is about 20\%-100\%, this high value due to that they just
used two-Gaussian to model the total H$\beta$ profile and assumed
the narrow H$\beta$ component is from NLRs. The flux ratio of
H$\beta$ to \oiii $\lambda$5007 from NLRs depend on the physics of
the low-density gas found in NLRs photoionized by AGN, which is
beyond the scope of this paper (e.g. Groves et al. 2006; Kewley et
al. 2006). During our fitting procedure of the H$\beta$ and \oiii
lines, we add a conservative constraint to the H$\beta$
contribution from NLRs, less than a half of \oiii $\lambda$5007
flux.

For 209 NLS1s in sample B, the second moment $\sigma_{\rm
H\beta}$ is calculated from the reconstructed H$\beta$ profiles
from BC and IC. We did a comparison between $\sigma_{\rm H\beta}$
and the FWHM of IC/BC. In Figure 4, we show $\sigma_{\rm H\beta}$
versus FWHM of IC and BC. The dashed line is the relation of $ \rm
FWHM = 2.35\times \sigma$ when a Gaussian profile is measured. We
find that there is a much stronger correlation between the
$\sigma_{\rm H\beta}$ versus FWHM of BC with respect to that for IC
(see Figure 4). If using the $\sigma_{\rm H\beta}$ to calculate
the SMBHs masses, the result depends much more on the BC FWHM, and less on the
narrower IC FWHM. We think that if the H$\beta$ profile from BLRs
can be fitted well by one Gaussian, for mass, there is no
difference for the usage of H$\beta$ FWHM and the $\sigma_{\rm
H\beta}$.

%Because NLS1s prefer the Lorentzian profile (Zhou et al. 2006;
%Veron-Cetty et al. 2001), the selection of $\sigma_{\rm H\beta}$ and
%FWHM of H$\beta$ can have an enormous effect on the mass
%calculation.

The existence of BC and IC in the H$\beta$ profile from BLRs have
been investigated by many people (Baldwin et al. 1998; Wills et al.,
1993; Brotherton et al. 1994; Rodriguez-Ardila et al. 2000; Leighly
2004; Dietrich et al. 2005), and it is suggested that BC and IC are
emitted from two distinct emission region. For 12 NLS1s, Dietrich et
al. (2005) found that $\rm FWHM_{BC}=3275\pm 800 \kms$ and $\rm
FWHM_{IC}=1200\pm 300 \kms$. For our 209 NLS1s in sample B, we also
found that $\rm FWHM_{BC}=4098\pm 1751 \kms$ and $\rm
FWHM_{IC}=1385\pm 492 \kms$. Therefore, it is clear that there
exists a BC in NLS1s that are typical for BLS1s, and the IC displays
typical H$\beta$ FWHM in NLS1s. The equivalent width (EW) of BC for
our 209 NLS1s in sample B is $19.4 \pm 8.6$\AA\, and $16.1 \pm
8.0$\AA\ for IC EW. There is no correlation between them.

\subsection{$\mbh - \sigma_{*}$ relation}
\begin{figure}
\begin{center}
\includegraphics[width=8cm]{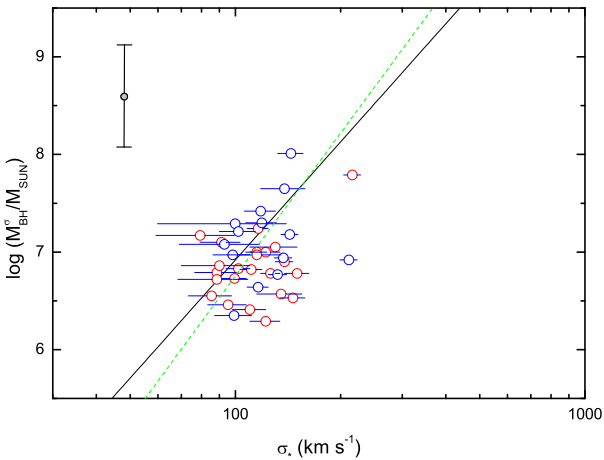}
\caption{The $\mbh - \sigma_{*}$ relation for 37 AGNs with
reliable measurements of $\sigma_{*}$. The blue and red circles
have the same meaning as in Figure 2. The solid line is the $\mbh -
\sigma_{*}$ relation of Tremaine et al. (2002). The Green dashed
line is the $\mbh - \sigma_{*}$ relation of Ferrarese \& Ford
(2005). The line in the left corner denotes the typical error in
mass calculation.}
\end{center}
\end{figure}

With FWHM from the H$\beta$ Lorentzian profile, Zhou et al. (2006)
use the $R_{\rm BLR} - L_{5100}$ relation of Kaspi et al. (2000)
and $f=3$ to calculate the mass. The updated $R_{\rm BLR} -
L_{5100}$ relation of Bentz et al. (2006) would lead to the mass of
Zhou et al. (2006) larger by 0.1-0.3 dex for NLS1s with mass less
than $10^7 \msun$ (see Figure 2b), which would place NLS1S close to
the Tremaine et al. $\mbh - \sigma_{*}$ relation when $\sigma_{*}$ is
adopted from $\sigma_{\rm [N II]}$ (see Figure 29 in Zhou et al.
2006). For a sample of 58 NLS1s selected from 11th edition of the
"Catalogue of Quasars and AGNs " (Veron-Cetty \& Veron 2003) and
SDSS DR3, by the $R_{\rm BLR} - L_{5100}$ relation of Kaspi et al.
(2005), Komossa \& Xu (2007) suggested that NLS1s do follow the
$\mbh - \sigma_{*}$ relation found in inactive galaxies after
excluding "blue outliers" (see Bian \& Zhao 2005).

Using the second moment of broad H$\beta$ profile, we find that
the SMBH mass would be larger by $\sim$ 0.5 dex with respect to that
from H$\beta$ FWHM in sample B with necessary two-Gaussian
fitting. The larger masses make NLS1s follow the Tremaine et al.
relation when the $\sigma_{\rm [O III]}^{core}$ is used as a
surrogate for the bulge velocity dispersion. In Figure 3, we find
that some AGN lie far below the $\mbh - \sigma_{*}$ relation and we
tried to determine if they are "blue outliers". No "blue outliers"
are found or it is impossible to measured the \oiii blueshift due
to the low S/N in \sii or \oii. We also use the \nii FWHM (Zhou et
al. 2006) as the $\sigma_*$, and find as similar result to that shown
in Figure 3, but with more scatter. The \nii FWHM measurement depends on the
H$\alpha$ profile fitting. And the \nii FWHM is consistent with
FWHM of \oiii core component, although the correlation is very
weak (Komossa \& Xu 2007).

In Figure 5, we plot \mbh from $\sigma_{\rm H\beta}$ versus
$\sigma_{*}$ for 37 NLS1s. For the $\sigma_{*}$ from SSP synthesis,
its uncertainty based on effective S/N at 4020\AA\ is typically
about: 24 \kms at S/N=5; 12 \kms at S/N=10; 8 \kms at S/N=15, where
the effective S/N is the S/N (measured between 4010 and 4060 \AA)
multiplied by the stellar fraction (Cid Fernandes et al. 2005; Bian
et al. 2007). The mass is calculated for AGNs that satisfy the first
two criteria in section 2. The blue circles denote AGN with
necessary two-Gaussian fits and the red circles denote AGN with
one-Gaussian fits, which is the same as that in Figure 2. For the
total 37 NLS1s in Figure 5, the mass is between $10^6-10^8 \msun$,
and the distribution of the mass deviation is $-0.24\pm 0.46$. For
14 AGN with two-Gaussian fits, the distribution of the mass
deviation is $-0.14\pm 0.50$, and for 23 AGN with one-Gaussian, the
distribution of the mass deviation is $-0.29\pm 0.43$. If these 23
NLS1s follow Tremaine et al. relation, we need $f=7.7$. The
$\sigma^{core}_{\rm [O III]}$ is slightly larger than $\sigma_{*}$
for these 37 NLS1s (see Botte et al. 2005). For about 3000 Seyfert 2
galaxies, Zhou et al. (2006) also found that $\sigma_{\rm [N
II]}=\sigma_{*}\times (2.62/2.35)$  and overestimate the $\sigma_*$
(also see Onken et al. 2004). Therefore, using the gas velocity
dispersion to trace $\sigma_{*}$ will place NLS1s close to the $\mbh
- \sigma_{*}$ relation and make the mass deviation smaller by 0.19
dex.

For the total 37 NLS1s, there exists marginal evidence that they
deviate from the $\mbh - \sigma_{*}$ relation found in inactive
galaxies. For 23 NLS1s with mass lower than $10^7 \msun$, the
deviation becomes much larger, up to $-0.46 \pm 0.40$ (also see
Figure 3), which is consistent with the result of Botte et al.
(2005, their figure 3), although it is not the case for the sample
of Greene \& Ho (2006) (also see Barth et al. 2005). If these 23
NLS1s follow the Tremaine et al. relation, we need $f=11.1$. It is
possible that the sample of Zhou et al. (2006) have some objects
that can't be classified by NLS1s. It is possible that there exists
true NLS1s with rapid growing small SMBH. The reliable measurements
of mass and the $\sigma_{*}$ are important for this kind of work.

\section{conclusions}
The second moment of H$\beta$ line from BLRs is calculated to derive
the SMBHs masses for a sample of 329 NLS1s selected from SDSS. The
main conclusions can be summarized as follows: (1) For objects with
necessary two-Gaussian fitting, the mean value of SMBH masses from
the H$\beta$ second moment is larger by about 0.5 dex with respect
to that from the H$\beta$ FWHM. (2) Using the narrow/core \oiii gas
velocity dispersion as a surrogate for the stellar velocity
dispersion, we find the new masses based on the H$\beta$ broad
emission line second moment bring them to the Tremaine et al.
relation; (3) The H$\beta$ second moment is more strongly correlated
with the BC FWHM rather than the IC FWHM. (4) Using the $\sigma_{*}$
measured from SSP synthesis, we find that, for NLS1s with masses
lower than $10^7 \msun$, they are marginally below the Tremaine et
al. relation considering the larger scatter in mass calculation. If
these 23 NLS1s follow Tremaine's relation, we need $f=11.1$.

\section*{ACKNOWLEDGMENTS}
We are very grateful to the anonymous referee for his/her
instructive comments. We are very grateful to Michael Brotherton for
his careful correction of our manuscript. We thank Luis C. Ho for
his useful comments, and thank discussions among people in IHEP AGN
group. This work has been supported by the NSFC ( Nos. 10403005,
10473005), the Science-Technology Key Foundation from Education
Department of P. R. China (No. 206053), and and the China
Postdoctoral Science Foundation (No. 20060400502). QSG would like to
acknowledge the financial supports from China Scholarship Council
(CSC) and the NSFC under grants 10221001 and 10633040. JMW is
supported by NSFC-10325313,10733010 and 10521001 and KJCX2-YW-T03,
respectively.

%\end{document}

\end{document}